\documentstyle[preprint,aps,prd]{revtex}

\begin{document}

\tightenlines

\title {Covariant perturbation theory and the Randall-Sundrum picture}

\author{Ezequiel Alvarez$^{1}$~\footnote{Electronic address:
    sequi@df.uba.ar} and Francisco D.\ 
  Mazzitelli$^{2}$~\footnote{Electronic address: fmazzi@df.uba.ar}}

\address{{\it $^1$ Instituto Balseiro, 
Centro At\' omico Bariloche\\ 8400 Bariloche, Argentina\\ 
    $^2$ Departamento de F\'\i sica, Facultad de Ciencias Exactas y
    Naturales\\ Universidad de Buenos Aires - Ciudad Universitaria, 
Pabell\'
    on I\\ 1428 Buenos Aires, Argentina}}

\maketitle
\begin{abstract}
The effective action for quantum fields on a $d$-dimensional spacetime
can be computed using a non local expansion in powers of the 
curvature. We show explicitly that, for conformal fields and up to quadratic
order  in the curvature, the non local effective action is equivalent 
to the
$d+1$ action for classical gravity in $AdS_{d+1}$ restricted to a
$d-1$-brane. This generalizes previous results about quantum 
corrections to the Newtonian potential and provides an alternative 
method for making local a non-local effective action. The 
equivalence can be easily understood by comparing the Kallen-Lehmann
decomposition of the classical propagator with the spectral representation
of the non local form factors in the quantum effective action.
\end{abstract}

\newpage

1. The analysis of the physical effects of quantum  fields on the 
background 
geometry requires the calculation of the effective action. This is 
a complicated object even for free fields. With the
exception of a few highly symmetric background metrics, it cannot be
computed exactly. Moreover, in order to study problems like
black hole evaporation or the physics of the early universe, it is 
necessary to compute the effective
action for an arbitrary metric,  
that should be fixed at the end by minimizing the effective action. 

A useful approach for the approximate computation of the effective
action is the so called covariant perturbation theory
\cite{barvi}. In this approach, that can be understood as a 
summation the Schwinger DeWitt expansion, the effective action is 
written in powers of curvatures. This approximation contains
non-local terms that include important physical information
like gravitational particle creation and the leading 
long distance quantum corrections to general relativity.

For conformal fields, in two spacetime dimensions the {\it quadratic} 
term
in the covariant
perturbation theory reproduces the (exact) Polyakov action. It is
possible to derive Hawking radiation from it \cite{russian}. 
In four dimensions
this has still not been done, and indeed it is a very difficult task
because Hawking radiation is contained in the {\it cubic} terms
of the expansion.

The four dimensional {\it quadratic} effective action has been used
to compute the leading long distance ${1\over r^3}$ corrections 
to the Newtonian potential \cite{dddm1}. 
\begin{equation}
V(r) = -{GM\over r}(1 - {(16\pi)^2 G B_4\over r^3})
\end{equation}
where the constant $B_4$ depends on the spin and number of quantum 
matter 
fields.
These corrections have
been computed by other methods a long time ago \cite{duff1}. 
For related works see \cite{newt}.

In a recent paper, Duff and Liu \cite{duff2} proved that
the same kind of corrections to the gravitational potential
do appear in the Randall-Sundrum brane-world proposal \cite{rs1}.
When a $3$-brane is inserted into $AdS_5$, and for classical matter
fields in the brane, the classical metric in five dimensions restricted
to the brane reproduces the classical Newtonian potential plus
the  ${1\over r^3}$ corrections. The coefficient of ${1\over r^3}$ 
that appears 
in this scenario coincides with the coefficient due to closed
loops of ${\cal N} = 4$ superconformal $U(N)$ Yang Mills theory
in the four dimensional theory. This is un tune with the
$AdS/CFT$ correspondence \cite{malda}.

In this paper we will extend the results of Ref. \cite{duff2}.
We will prove that, up to quadratic order in the curvature and
for free conformal fields, the non local $d$ dimensional effective 
action coincides with the restriction of the gravitational action
in $AdS_{d+1}$ to a $d-1$-brane. The results are valid 
for $d>2$.
We stress that we are not trying to check the consistency between
the AdS/CFT and the brane-world relations, as in
Ref. \cite{duff2}. Our aim is
to provide an alternative representation for the non local
$d$ dimensional effective action.

\bigskip

2. For a scalar field in curved spacetimes the effective action is 
given by $
\Gamma = {1\over 2}\ln\det {{\cal O}\over\mu^2}
$
where ${\cal O} = - g^{\mu\nu}\nabla_{\mu}\nabla_{\nu} + m^2 + \xi R$ 
is the operator
of the classical field equation, $\mu$ is an arbitrary parameter
with dimensions of mass,
$m$ is the mass of the scalar field,
 and $\xi$ is the coupling to the scalar curvature.
The conformal coupling in $d$ dimensions is $\xi = \xi_{c} = 
\frac{1}{4} \frac{d-2}{d-1}$

Using heat kernel techniques \cite{heat}
it is possible to obtain the Schwinger DeWitt expansion for the 
effective action
\begin{equation}
\Gamma = -{1\over 2 (4\pi)^{d/2}}
\int_0^\infty {ds\over s^{1+{d\over 2}}}e^{-m^2 s}\sum_{l\geq 0}
{(-s)^l\over l!} \int d^dx\sqrt g a_l(x)
\label{sdw}
\end{equation}
The Schwinger DeWitt coefficients $a_l$ are functions of the curvature
and its covariant derivatives. When integrating out term by term the
expression above, an expansion in inverse powers of the mass is
obtained. The expansion is valid for slowly varying metrics
that satisfy ${\cal R}\ll m^2$ (${\cal R}$ denotes components of
the curvature tensor). 
The expansion is local, and adequate for the analysis of the 
divergences of the
theory, which are contained
in the terms with $l$ less or equal to the integer part of
$\frac {d}{2}$. However, it misses very important
physical effects (like particle creation), and it is not adequate for 
massless quantum fields.

It is possible to perform a partial summation of the Schwinger DeWitt
expansion by keeping terms up to a given order in the curvature. The
idea was introduced in Ref. \cite{gospel} and further developed in Ref.
\cite{barvi,avra}. The effective action 
for a massless scalar field
in $d$ spacetime dimensions,
up to quadratic order in the curvature, can be written as \cite{avra}
\begin{equation}
\Gamma = \Gamma_{local} + \Gamma_{nonloc}
\label{accion}\end{equation}
where
\begin{equation}
\Gamma_{local} =  \int d^d x \sqrt{-g} \, \left[ 
- \Lambda_{(d)} + M^{d-2}_{(d)} 
\, R_{(d)} - \sum_{l=0}^{k} \left( \alpha^{(l)} R_{(d)} \Box^l  R_{(d)} 
+ 
\beta^{(l)} R_{\mu\nu (d)} \Box^l R^{\mu\nu}_{(d)} \right) \right]
\label{local}\end{equation}
and
\begin{equation}
\Gamma_{nonloc} =  -\alpha \, \int d^d x \sqrt{-g} 
\left( a \, R_{(d)} \, f(\Box) \, R_{(d)} + b \, R_{\mu \nu (d)}\, 
f(\Box) 
\, R^{\mu \nu}_{(d)} \right).\label {nonlocal}\end{equation}
Here $g_{\mu\nu}$, $x^{\mu}$, $M_{(d)}$ and 
$\Lambda_{(d)}$ are the d-dimensional metric,
coordinates, Planck mass 
($M_{(4)}^2 = 1/16 \pi G_{Newton}$) and cosmological constant, 
$R_{(d)}$ and $R_{\mu\nu (d)}$ the 
Ricci tensor and scalar respectively, 
\begin{eqnarray}
\alpha &=& (4\pi)^{-d/2} \, \frac{\sqrt{\pi}}{8\Gamma( (d-1)/2 )},
\nonumber \\ 
\nonumber
a &=& ( \xi - \xi _{c} )^{2} - \frac{d}{8 (d-1)^{2} \, (d+1)} , \\ 
\nonumber
b &=& \frac {1}{2( d^{2}-1)},
\label{ayb} 
\end{eqnarray}
and 
\begin{eqnarray} f(\Box) = \left\{ \begin{array}{ll}
  (-1)^{\frac{d}{2}} ( \frac{- \Box}{4} ) ^{d/2-2} \, 
\ln ( \frac{- \Box}{4\mu ^{2}}) 
& \mbox{d even} \\
(-1)^{\frac{d-1}{2}} \pi ( \frac{- \Box}{4}) ^{d/2-2} & 
\mbox{d odd.} \end{array} \right. \nonumber 
\end{eqnarray}
The summation in Eq.[\ref{local}] runs up to $k$, the integer 
part of $d/2-2$. These terms
are needed to renormalize the theory. After renormalization, 
the coefficients
$\alpha^{(l)},\beta^{(l)}$ and $\Lambda_{(d)}$ might take arbitrary values. 
For simplicity 
we will take $\Lambda_{(d)} = 0$ in what follows.

The results for the effective action can be extended for fields of 
arbitrary spin \cite{avra}. For example, for a massless Dirac field
in four dimensions, the effective action is six times the result
for a conformally coupled scalar field \cite{phdalvit}.

For $g_{\mu\nu} = \eta_{\mu\nu} + h_{\mu\nu}$,
and in the harmonic gauge (i.e.
$h_{\mu\nu} , ^{\nu} = \frac{1}{2} h_{\alpha} \, ^{\alpha} , _{\mu}$ ), 
Eq. [\ref{accion}] 
can be rewritten as
\begin{eqnarray}
\Gamma ^{(2)} = \frac{-1}{4} \int d^{d}x \; h_{\mu\nu} 
\triangle^{-1\mu\nu} 
\;_{\rho\sigma} h^{\rho\sigma} 
\label{orden2},
\end{eqnarray}
where
\begin{eqnarray}
\triangle^{-1\mu\nu} \, _{\rho\sigma} = - M^{d-2}_{(d)} \, \Box \,  
\left( \eta^{\mu} _{\rho} \eta^{\nu} _{\sigma} - \frac{1}{2} \eta^{\mu\nu} 
\eta_{\rho\sigma} \right) + \alpha \Box^2 \, f(\Box) \left( b \, 
\eta^{\mu} _{\rho} 
\eta^{\nu} _{\sigma} + a \, \eta^{\mu\nu} \eta_{\rho\sigma} \right) 
\nonumber \\+ \sum_{l=0}^k \Box^{l+2} \left( \alpha ^{(l)}  
\eta^{\mu\nu} 
\eta_{\rho\sigma} + \beta ^{(l)}  \eta^{\mu} _{\rho} \eta^{\nu} 
_{\sigma} 
\right) .\label{inverse} \end{eqnarray}

If we add to the theory classical matter described by an energy 
momentum tensor
$T^{\mu\nu}$, the
spacetime metric  satisfies 
\begin{equation}
h_{\mu\nu} = \triangle ^{\alpha\beta} \, _{\mu\nu} T_{\alpha\beta} .
\label{ec. de mov.}
\end{equation}
In the low-energy approximation the quantum correction 
in Eq.[\ref{inverse}] can be treated as a small perturbation.  
Using that the inverse of
$A  \eta^{\mu\nu} \eta_{\rho\sigma} + B \eta^{\mu} _{\rho} 
\eta^{\nu}_ 
{\sigma}\,\,\,$ is given by 
$\frac{-A}{B(B+A \cdot d)} \eta^{\alpha\beta} \eta_{\mu\nu} + 
\frac{1}{B} \eta^{\alpha} _{\mu} \eta^{\beta} _{\nu},$  
it is straightforward 
to check that,
up to 
leading order,
\begin{eqnarray}
\triangle^{\alpha\beta}\, _{\mu\nu} = \frac{-1}{M^{d-2}_{(d)} \, \Box} 
\,   \left( \eta^{\alpha} _{\mu} \eta^{\beta} _{\nu} - 
\frac{1}{d-2} \eta^{\alpha\beta} 
\eta_{\mu\nu} \right) - \frac{\alpha}{2(d^{2}-1)\,M^{2(d-2)}_{(d)} } \, 
f(\Box) \nonumber\\ \left(  \eta^{\alpha} _{\mu} \eta^{\beta} _{\nu} - 
\left[ \frac{1}{d-1} - (\xi - \xi_{c})^{2} \, 8 \frac{d^{2} - 
1}{(d-2)^{2}} 
\right] \eta^{\alpha\beta} \eta_{\mu\nu} \right) + \sum_{j=0}^{\tilde k} 
g_{j}^{(1)} 
\Box^j  \eta^{\alpha} _{\mu} \eta^{\beta} _{\nu} + g_{j}^{(2)} \Box^j 
\eta^{\alpha\beta} \eta_{\mu\nu} .  
\label{propagador}
\end{eqnarray}
The constants $g_{j}^{(i)}$  depend on $d$, $M_{(d)}$, 
$\alpha^{(l)}$ and 
$\beta^{(l)}$.  The summation here and in what follows is only for 
$d\geq5$; $\tilde k$ is equal to $k$ when $d$ is odd and to $k-1$ when 
$d$ is even.

The meaning of 
Eq. [\ref{propagador}] is very simple: the first term 
corresponds to the classical propagation while the second contains 
the quantum corrections and is easily traced back to the non-local part 
of the action, 
Eq. [\ref{nonlocal}].  The analytic terms (proportional to $\Box^j$), 
will not contribute in the large distance/low-energy  limit (see 
below).

\bigskip

3. We will now prove that a propagator similar to Eq. 
[\ref{propagador}] 
describes the classical propagation on a brane inserted into 
$AdS_{d+1}$\cite{giddins}. 
If the d-dimensional space-time is thought as a $d-1$ brane in a 
(d+1)-dimensional theory then the classical action reads
\begin{eqnarray}
\int d^{d+1} X \,\sqrt{-G} \left( M_{(d+1)}^{d-1} R_{(d+1)} - 
\Lambda_{(d+1)} + 
{\cal L}_{matter} \right) - \int d^d x \,\sqrt{-g} \, \tau.
\end{eqnarray}
Here  $X^I = (y,x^\rho )$
are the (d+1)-dimensional coordinates,
$G_{IJ}$ the metric in (d+1) dimensions, and $\tau$ the brane tension.  
The ${\cal L}_{matter}$ term may include a matter source in the brane 
as 
well as in the bulk. The brane geometry is chosen such that $y$ is the 
coordinate in the bulk and $x^{\rho}$ are coordinates along the brane 
(which is located at $y=0$), then small fluctuations to the metric are 
represented by
\begin{eqnarray}
ds^2 = dy^2 + e^{-2|y| / L} \left( \eta_{\mu\nu} + h_{\mu\nu} 
(x^{\rho},y) 
\right) dx^\mu dx^\nu,
\end{eqnarray}
where $L=\sqrt{-d(d-1) M_{(d+1)} ^{\,d-1} / \Lambda_{(d+1)}}$. 

We are only interested in $h_{\mu\nu} (x^\rho , y=0)$ 
when the matter source is located on the brane. 
In this situation, it has been shown that the effective propagator
on the brane is given by \cite{giddins} 
\begin{eqnarray}\triangle^{\alpha\beta} \, _{\mu\nu} = 
-\frac{d-2}{LM_{(d+1)}^{d-1}} \cdot \frac{1}{\Box} 
\left(\eta^{\alpha}_{\mu} 
\eta^{\beta}_{\nu} - \frac{1}{d-2} 
\eta^{\alpha\beta} \eta_{\mu\nu} \right) -  
\frac{1}{M_{(d+1)}^{d-1}} \triangle_{KK} (\sqrt{-\Box}) 
\left(\eta^{\alpha}_{\mu} \eta^{\beta}_{\nu} - \frac{1}{d-1} 
\eta^{\alpha\beta} \eta_{\mu\nu} \right), 
\label{propagador brana}
\end{eqnarray}
where 
\begin{eqnarray}\triangle _{KK} (\sqrt{-\Box}) = 
\frac{-1}{\sqrt{-\Box}} 
\frac{K_{d/2-2} (\sqrt{-\Box} L)}{K_{d/2-1}(\sqrt{-\Box} L)}\,\, .
\label{KK}
\end{eqnarray}
Again Eq.[\ref{propagador brana}] has a simple interpretation: the first
term describes the zero mode graviton localized on the brane, while the 
second term
corresponds to the continuum Kaluza Klein graviton modes.

At large distances, corresponding to $\sqrt{-\Box} L \ll 1$, 
Eq.[\ref{KK}] 
can be expanded to give,
up to the first term non analytic-in-$\Box$,
\begin{eqnarray} {1\over L}\triangle_{KK} (\sqrt{-\Box}) 
\approx \left\{ \begin{array}{ll}   \sum_{l=0}^{\tilde k} c^{(e)}_l 
(\Box L^2)^l + c^{(e)}_{d/2-2} (-1)^{\frac{d}{2}} (\frac{-\Box}{4} 
L^2)^{d/2-2} \, \ln (-\Box L^2) & 
\mbox{d even} \\  \sum_{l=0}^{\tilde k} c^{(o)}_l (\Box L^2)^l 
+ \pi c^{(o)}_{d/2-2} (-1)^{\frac{d-1}{2}}(\frac{-\Box}{4} L^2)^{d/2-2} & 
\mbox{d odd.} 
\end{array} 
\right. 
\label{KK aprox}
\end{eqnarray}
The coefficients $c^{(i)}_l$ can be easily obtained, but we will not 
need 
the explicit expression in what follows.

Now we compare Eqs.[\ref{propagador brana}] and [\ref{propagador}].
The classical terms in both propagators coincide if we
choose the coupling constants such that
$\frac{d-2}{LM_{(d+1)}^{d-1}}=\frac{1}{M^{d-2}_{(d)}}$.
In order to have agreement between the leading non analytic terms,
the coupling must be conformal, i.e. $\xi = \xi_c$. Moreover,
we must have $ c_{d/2-2}^{(e,o)} \cdot L^{d-3} / 
M_{(d+1)}^{d-1} = \alpha (2(d^2 -1) M_{(d)}^{2(d-2)})^{-1}$. 
These equations relate the values 
of the $d+1$  cosmological constant and Planck mass with Planck mass in 
$d$ dimensions.
Had we considered a  different free field content 
on the brane ($N_s$ fields of spin $s$)  the only difference would have 
been a different relation between the values of these parameters.

It is not necessary to require 
agreement between the terms analytic in $\Box$ since, as shown below,
they are not relevant in the low-energy limit. However, it is 
worth noting that with the choice 
$g_j^{(2)} = -\frac{1}{d-1}g_j^{(1)}$ and 
$g_j^{(1)}=-c^{(e,o)}_j \cdot L^{2j+1} / M_{(d+1)}^{d-1}$
the analytic terms also coincide.
This 
would imply particular values for the constants 
$\alpha^{(l)}$
and $\beta^{(l)}$ in the non local effective action 
Eq.[\ref{nonlocal}], all of them determined by
$M_{(d+1)}$ and $\Lambda_{(d+1)}$. 

We will now show that analytic terms 
are not relevant in  the low-energy limit. 
To illustrate this point we  compute
the quantum corrections to the $d$-dimensional Newtonian potential, 
$-\frac{1}{2} h^{00} (x)$. We assume a classical mass $M$ fixed at the 
origin of 
coordinates, namely
$T^{\mu\nu} (x) = \delta^{\mu}_{0} \delta^{\nu}_{0} 
\, M
\,\delta^{(d-1)} (\vec{x})$.
Here $\vec{x}$ are the space coordinates, 
$\vec{x}=(x_1 , x_2 , \, ...\, ,x_{d-1} )$.  With this in mind, 
using Eqs.[\ref{ec. de mov.}] and [\ref{propagador}], 
the quantum corrected Newtonian potential reads
\footnote{Note that for $d=3$ spacetime is flat outside
matter, hence there is no gravitational force. The term
proportional to $B_3$ comes from the quantum correction.}
\begin{eqnarray}
V(r) =\frac{-1}{2} h^{00} (r) = \left\{ \begin{array}{ll}  
B_3 \frac{M}
{M_{(3)}^2} \cdot \frac{1}{r} & 
d=3 \\ A_d \frac{M}
{M_{(d)}^{d-2}}\frac{1}{r^{d-3}} + 
B_d \frac{M}
{M_{(d)}^{2(d-2)}} \cdot \frac{1}{r^{2d-5}} 
-{M\over 2} 
\sum_{j=0}^{\tilde k} (g_{j}^{(1)} 
\Box^j  + g_{j}^{(2)} \Box^j )\delta^{(d-1)} (\vec{x}) &
d\geq 4 \end{array} \right. \nonumber
\end{eqnarray}
where $A_d$ and $B_d$ are constants and $r=| \vec{x} |$. 
As anticipated, the analytic terms proportional to $\Box^j$
produce quantum corrections localized at the origin. They are
therefore irrelevant at large distances. In four dimensions,
the Newtonian potential reads
\begin{equation}
V(r) = -\frac {GM}{r}\left[ 1+\frac{G}{45 \pi r^2}\left (
1  + 45 (\xi -\frac{1}{6})^2\right )\right ]\,\, ,
\end{equation}
and agrees with previous results for $\xi = 1/6$
\cite{duff2,phdalvit}. If we consider $N_0$ scalar fields and
$N_{1/2}$ Dirac fields, the Newtonian potential becomes
\begin{equation}
V(r) = -\frac {GM}{r}
\left\{ 1+\frac{G}{45 \pi r^2}\left [
N_0 \left ( 1  + 45 (\xi -\frac{1}{6})^2\right )
+ 6 N_{1/2}\right ]\right \}\,\, .
\end{equation}
\bigskip

4. Non local effective actions have been previously localized through the 
introduction 
of  auxiliary fields. 
For example, in two dimensions, Polyakov's action
\begin{equation}
S_P =- {1\over 96\pi}\int d^2x\sqrt{-g}R{1\over \Box}R
\end{equation}
can be made local by introducing an auxiliary field $\psi$
and the local action
\begin{equation}
S_{local} = -{1\over 96\pi}\int d^2x\sqrt{-g}(-\psi
\Box
\psi + 2\psi R)
\end{equation}
On shell for the auxiliary field, both actions $S_P$ and $S_{local}$
are equivalent.

In  four dimensions, the effective action that reproduces 
the conformal anomaly is the so called Reigert's action \cite{reig}.
The non local part of the Reigert's action is, schematically,
\begin{equation}
S_R = \int d^4x {\cal R}^2 
{1\over \triangle_4}
{\cal R}^2
\end{equation}
where ${\cal R}$ denotes components of the Riemann tensor and
$\triangle_4$
is the fourth order operator 
$\triangle_4 = \Box^2-2R^{\mu\nu}\nabla_{\mu}\nabla_{\nu}
+ \frac{2}{3}R\Box - \frac{1}{3}\nabla^{\mu}R\nabla_{\mu}\,\, .$
Reigert's action can be made local 
\cite{reig,shap} by the introduction of 
auxiliary scalar fields 
\begin{equation}
S_{local} = \int d^4x \left (-\psi\triangle_4\psi + 2 \psi {\cal R}^2
\right ) 
\end{equation}

The localization based in the introduction of auxiliary fields
works only when the form factors in the non local
effective action are the inverse of polynomials in $\Box$ and 
$\nabla_{\mu}$.
Here the form factors does not satisfy this property. 
An extra dimension
is needed to make local the action. 
The mathematical reason for this can be 
understood as follows. The non analytic form factors can be represented
in the form of spectral integrals \cite{barvi}. For example,
in three and four dimensions
the form factors can be written as
\begin{eqnarray}
\ln\left (\frac{-\Box}{\mu^2}\right ) &=& \int_0^{\infty}d\lambda
\left(\frac {1}{\lambda - \Box} - \frac {1}{\lambda + \mu^2}\right )\nonumber\\
(-\Box)^{-1/2} &=& \frac{2}{\pi}
 \int_0^{\infty}d\lambda
\left(\frac {1}{\lambda^2 - \Box}\right )
\label{spec}
\end{eqnarray} 
Similar expressions can be found for other dimensions.
Note that the non analytic functions of $\Box$ are written as 
integrals that involve massive propagators ($\frac {1}{\lambda - \Box}$
or $\frac {1}{\lambda^2 - \Box}$).

On the other hand, the restriction of a massless $d+1$ propagator
on a $d-1$ brane also admits an analogous representation
\cite{warped}. Indeed, let us consider the metric
$ds^2 = dy^2+w^2(y)g_{\mu\nu}(x)dx^{\mu}dx^{\nu}$. The
D'Alambertian operator can be written as
\begin{equation}
\Box_{d+1} = \frac {\Box}{w^2} + \frac {\partial^2}{\partial
y^2} + d\frac{w'}{w}\frac{\partial}{\partial y}\equiv
\frac {\Box}{w^2} + \Box_y
\end{equation}
where $\Box$ is the $d$-dimensional D'Alambertian associated
to $g_{\mu\nu}$.

We introduce the eigenfunctions $\theta^{(i)}_{\lambda}(y)$, that 
satisfy $\Box_y \theta^{(i)}_{\lambda} =  -
\frac{\lambda}{w^2}\theta^{(i)}_{\lambda}$. It can be 
easily shown \cite{warped} that the massless propagator
$\triangle = \frac {1}{\Box_{d+1}}$, restricted to a 
fixed slice $y=const$, admits the following representation
\begin{equation}
\triangle(x,y,x',y) = \sum_{i,\lambda}
\vert\theta^{(i)}_{\lambda}(y)\vert ^2\frac{1}{\Box - \lambda}
\label{lehman}
\end{equation}
This is analogous to the Kallen-Lehman decomposition in
quantum field theory, with a weight function
$\mu(\lambda , y)=  \sum_{i}
\vert\theta^{(i)}_{\lambda}(y)\vert ^2$.

The similarity between the form factors in the non local quantum
effective action and the restriction of the classical propagator
on a brane is now clear (compare Eqs. [\ref{spec}] and [\ref{lehman}]).
Roughly speaking, in this paper we have shown that the weight function
in $AdS_{d+1}$ spacetime 
reproduces the spectral representation of the $d$-dimensional
form factor for conformal fields.
It is 
possible that, by taking a different metric in the bulk, one could
reproduce the non local effective action for non conformal fields.
Alternatively, a different quantum field theory on the brane
could reproduce the $AdS_{d+1}$ propagator beyond leading order.

The equivalence shown in this letter could be useful as a tool
for computations of the effects of quantum fields on the
spacetime metric, since it may be technically more easy
to work with an extra dimension than with a non local 
effective action.

\acknowledgments This work was supported by Universidad de Buenos 
Aires,
CONICET (Argentina) and  CNEA (Argentina). We would like to
thank G. Giribet and J. Russo for useful conversations.
 
\samepage{

}
\end{document}